\documentclass[12pt]{article}

\input psfig.sty
\usepackage{amsfonts}
\usepackage{amssymb}
\newtheorem{thm}{Theorem}
\def\pf{\medbreak\noindent{\bf Proof:}\enspace}

\def\iff{\Longleftrightarrow}

\def\T{\hbox{Tr}}

\title{A One-Dimensional Model for Many-Electron Atoms
 in Extremely Strong Magnetic Fields:  Maximum Negative Ionization}

\author{Raymond Brummelhuis \\
Department of Mathematics \\  Universit\'e de Reims, 
\\ F-51687 Reims, France \\ 
{\normalsize raymond.brummelhuis@univ-reims.fr}
\and Mary Beth Ruskai \thanks{supported by National Science
Foundation Grant DMS-94-08903} \\ Department of
Mathematics \\ University of Massachusetts  Lowell \\ Lowell,
MA  01854 USA \\ {\normalsize bruskai@cs.uml.edu} }

\date{18 December 1998}

\begin{document}

\maketitle

\begin{abstract}

We consider a one-dimensional model for many-electron atoms in
strong magnetic fields in which the Coulomb potential and
interactions are replaced by one-dimensional regularizations
associated with the lowest Landau level.  For this model we
show that the maximum number of electrons $N_{\max}$ satisfies a
bound of the form $N_{\max} < 2Z+1 + c \sqrt{B}$ where $Z$ denotes
the charge of the nucleus, $B$ the field strength and $c$ is a
constant. We follow Lieb's strategy in which convexity plays a
critical role.  For the case $N=2$ with fractional
nuclear charge, we also discuss the critical
value $Z_c$ at which the nuclear charge becomes too weak to
bind two electrons.

\end{abstract}

\pagebreak

\section {Introduction}

It is well-known that systems in strong magnetic fields behave 
like systems in one-dimension, i.e., a strong magnetic field  
confines the particles to Landau orbits orthogonal to the
field, leaving only their behavior in the direction of the field
subject to significant influence by a static potential.
Therefore, a better understanding of one-dimensional systems
is essential to understanding the behavior of systems in strong
magnetic fields.  Although many one-dimensional systems, including that of
 a hydrogen atom with a single electron \cite{AHS2,HR}, have been thoroughly
studied, relatively little is known about multi-particle atoms
confined to one dimension. 

In this paper we study the question of bounds on the maximum excess
negative charge using one-dimensional models of many-electron atoms.
Because our goal is an understanding of the behavior in
one-dimension, we do not deal with  the question of
accuracy of our one-dimensional models as approximations to, or
reductions from, real 3-dimensional atoms.  However, we sketch
such a reduction as motivation for the models considered.   

There is some question as to the proper replacement
for the Coulomb potential in one dimension \cite{HR}.  The potential
$V(x) = 1/|x|$ is so singular that the one-dimensional Hamiltonian
$ - \Delta - 1/|x|$ is not even essentially self-adjoint.  Fortunately,
an electron in a Landau orbit is attracted to a nucleus with
charge $Z$, not by a potential of the form $-Z/|x|$, but by a regularized
potential which is finite at the nucleus.  However, the
corresponding regularization of the electron-electron interaction
is more complicated unless both electrons have zero angular
momentum in the direction of the field.  In that case, the
regularized interaction has a simple form which is the basis for
our model.

There is an extensive literature on atoms in magnetic fields.  
Interest in atoms in extremely strong fields, which began in the 1970's
after the discovery of pulsars, has recently been renewed in the
1990's, culminating in the comprehensive work of 
Lieb, Solovej and Yngvason  (LSY) \cite{LSY1,LSY2, LSY3, Y}.
For a discussion of early work on approximations for atoms in extremely
strong magnetic fields we refer the reader to 
the insightful paper of Rau, Mueller, and Spruch \cite {RMS}
and to the review by Ruderman\cite{Rud}.
References to later work are given in the
introduction to  \cite{LSY2} and a summary of the work
of  LSY \cite{LSY2, LSY3} is
given in \cite{LSY1, Y}.  Rigorous work on atoms in magnetic
fields began with the work of Avron, Herbst and Simon (AHS)
\cite{AHS1, AHS2, AHS3}.  LSY \cite{LSY1,LSY2, LSY3, Y}
 not only analyzed extensions of Thomas-Fermi
theory in 5 distinct regions, but showed that
these regions suffice to give the correct asymptotics for the exact
Hamiltonian.  In particular, they showed that the maximum number
of electrons $N_{\max}(Z,B)$ which can be bound to a nucleus of
charge $Z$ in a constant magnetic field of strength $B$ 
satisfies $\liminf N_{\max}(Z,B)/Z \geq 2$ as $Z \rightarrow \infty$
and $B/Z^3 \rightarrow \infty.$  This should be compared with
asymptotic
neutrality \cite{LSST, FS, SSS}, i.e., 
$\lim_{Z \rightarrow \infty} N_{\max}(Z,0)/Z = 1$, for
atoms without magnetic fields.  We hope that the analysis of the
simple model in this paper is a modest first step toward a better
understanding of the mechanism by which extremely strong magnetic fields
bind an ``extra'' $Z$ electrons, as well as the conjectured converse
$N_{\max}(Z,B) \leq 2Z.$

The full 3-dimensional Pauli Hamiltonian for an N-electron atom with 
nuclear charge $Z$ in a constant magnetic field of strength $B$,
acting on ${\cal H}^N$, the n-fold tensor product of 
$L^2({\bf R}^3) \otimes {\bf C}^2$, is
\begin{eqnarray}
  {\bf H}(N,Z,B,\alpha) = \sum_{j=1}^N \left[ |{\bf P}_j + {\bf A}|^2 + 
    {\bf \sigma}_j \cdot {\bf B} - Z/|{\bf r}_j| \right] +
     \sum_{j<k} \alpha/|{\bf r}_j - {\bf r}_k |
\end{eqnarray}
where ${\bf A}$ is a vector potential such that 
  ${\bf \nabla}  \times  {\bf A} = {\bf B}$ and $\alpha$ is a coupling
constant introduced for convenience in discussing scaling.  
If the spin-coupling term is omitted, it often suffices to consider
the corresponding scalar Hamiltonian, which we denote $H(N,Z,B,\alpha)$,
as an operator acting only on $[L^2({\bf R}^3)]^N$ or
the ``space"  portion of ${\cal H}^N$. We will choose
our coordinate system so that the field  ${\bf B} = (B,0,0)$ is in the
 x-direction and the gauge so that $2{\bf A} = {\bf B} \times {\bf r}$.
The Hamiltonian (1) satisfies the scaling relation
   ${\bf H}(N,Z,B,1) = B {\bf H}(N,Z B^{-1/2},1,B^{-1/2})$, 
i.e., we can scale out 
the field strength by replacing the nuclear charge $Z$ by $Z B^{-1/2}$
and reducing the electron-electron interaction by $B^{-1/2}$.
Alternatively, we could have included the electron charge unit  $e$
explicitly and replaced $e$ by $e B^{-1/4}$.

We let $E_0(N,Z,B,\alpha)$ denote the infimum of the spectrum of
the scalar Hamiltonian $H(N,Z,B,\alpha)$ defined above.
It is well-known that the spectrum of the
``free'' Hamiltonian  $|{\bf P}_j + {\bf A}|^2$ is $[B,\infty]$,
that the spectrum of $|{\bf P}_j + {\bf A}|^2 + {\bf \sigma}_j \cdot {\bf B}
  = [{\bf \sigma}_j \cdot ({\bf P}_j + {\bf A})]^2 $ is $[0,\infty)$
and that the lowest Landau level has energy $B$ with infinite
degeneracy indexed by $m = 0, 1, 2, 3, .... $ corresponding to
angular momentum $-m$ quantized in the field direction.
Therefore, the continuous spectrum of 
  $H(N,Z,B,\alpha)$ is $[ B + E_0(N-1,Z,B,\alpha), \infty)$, i.e.,
the continuum begins at $B$ plus the ground state energy for $N-1$ electrons.
Thus the question of whether or not $H(N,Z,B,\alpha)$ has a bound state
is determined by whether or not some test function $\Psi$ satisfies
\begin{eqnarray}
  \langle \Psi, H(N,Z,B,\alpha) \Psi \rangle  <  
    [B + E_0(N-1,Z,B,\alpha)] \| \Psi \|^2.
\end{eqnarray}
For the full Hamiltonian  ${\bf H}(N,Z,B,\alpha)$, the continuous
spectrum is $[E_0(N-1,Z,B,~\alpha), \infty)$ which implies that 
${\bf H}(N,Z,B,\alpha)$ has a bound state if and only if there is a
${\bf \Psi} \in {\cal H}^N$ for which
\begin{eqnarray}
  \langle {\bf \Psi}, {\bf H}(N,Z,B,\alpha) {\bf \Psi} \rangle  
    <  E_0(N-1,Z,B,\alpha)  \| {\bf \Psi} \|^2.
\end{eqnarray}
For the remainder of this paper we will omit explicit consideration
of spin [although we will be able to draw some conclusions indirectly.
See the remark after equation (\ref{varequivspin}).]  
Our methods cannot handle the   ${\bf \sigma}\cdot {\bf B}$
term explicitly and the inclusion of spin in the wave function does not
affect the remaining results in any essential way.

\begin{figure}
\vskip4cm
\hskip4cm\psfig{figure=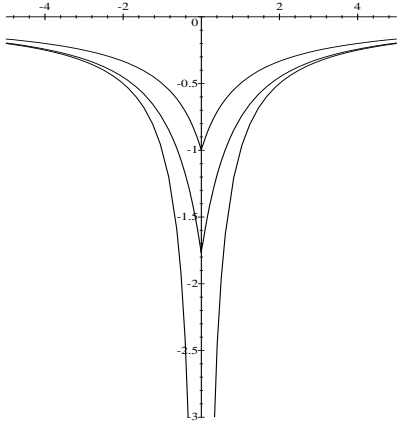,height=1.2in,width=1.2in}
\vskip-0.5cm
\caption{Comparison of the potentials (reading from the top down)
$-\frac{1}{|x|+1}$, $-V_0(x)$, and $-\frac{1}{|x|}$}
\label{fig:V0}
\end{figure}

We now consider the  the 3-dimensional energy minimization
problem restricted to functions whose behavior orthogonal to the
field is described entirely by product functions in which all electrons
are confined to the lowest Landau level, i.e we restrict to
 N-electron functions $\Psi$ of the form 
       $\Psi_{m_1 \ldots m_N} = 
      \Phi(x_1 \ldots x_n) \prod_{k=1}^N  \gamma_{m_k}^B(r_k,\theta_k)$
where  
  $$\gamma_m^B(r,\theta) = 
        [\pi m!]^{-1/2}B^{(m+1)/2} r^m e^{-Br^2/2}e^{-i m \theta}$$
denotes the Landau level with energy $B$ and angular momentum $-m.$
(Note that we are using cylindrical coordinates $(x,r,\theta)$ with
$r = \sqrt{y^2 + z^2}$ so that $|{\bf r}| = \sqrt{x^2 + r^2}$.)
 Such expectations satisfy
\begin{eqnarray} 
\lefteqn{\langle \Psi_{m_1 \ldots m_N}, H(N,Z,B,1) \Psi_{m_1 \ldots m_N} 
    \rangle} ~~~ \nonumber \\
  & = &   \langle \Psi_{m_1 \ldots m_N},  
       \widehat{H}(N,Z,B,1) \Psi_{m_1 \ldots m_N} \rangle  + NB   \\
   & = &  B \langle \Psi_{m_1 \ldots m_N},  
         \widehat{H}(N,ZB^{-1/2},1, B^{-1/2}) \Psi_{m_1 \ldots m_N} \rangle 
                       + NB  \nonumber
\end{eqnarray}
where 
\begin{eqnarray}
\widehat{H}(N,Z,B,\alpha) = \sum_{j=1}^N 
   \left[ - |P_j^x|^2  - Z/|{\bf r}_j| \right] +
     \sum_{j<k} \alpha/|{\bf r}_j - {\bf r}_k |.
\end{eqnarray}
For each fixed choice of $m_1 \ldots m_N$ the minimization problem
can be reduced to a one-dimensional problem for the Hamiltonian
$\widehat{H}_x^{m_1 \ldots m_N}(N,Z,B,\alpha)$ in which the kinetic
energy has the usual $-\frac{d^2}{dx^2}$ form and the
potentials $1/|{\bf r}_j|$ and $1/|{\bf r}_j - {\bf r}_k|$ 
are replaced by regularized potentials denoted by $V_{m_j}^B(x_j)$
 and    $W_{m_j,m_k}^B(|x_j - x_k|)$ respectively.
\begin{eqnarray}
  V_m^B(x) & = & \int\!\int |\gamma_{m}|^2 / |{\bf r}|dy dz \nonumber \\
     & = & \frac {B^{m+1}}{m!} \int_0^\infty 
        \frac{r^{2m} e^{-Br^2} }{\sqrt{x^2 + r^2}} r dr  \nonumber \\
     & = &  [ m!]^{-1} \int_0^\infty 
          \frac{u^m e^{-u} }{\sqrt{x^2 + u/B}} du  \label{vmdef} \\
    & = & \frac {2B^{m+1}}{m!} e^{Bx^2}\int_{|x|}^\infty  
             (t^2 - x^2)^m e^{-Bt^2} dt  \nonumber
\end{eqnarray}
which satisfies the scaling relation  
$V_m^B(x) = \sqrt{B}~ V_m^1(\sqrt{B}x)$ as one would expect from
the scaling properties of $H(N,Z,B,\alpha)$.
$W_{m_j,m_k}^B$ is defined analogously; we postpone discussion
of its explicit form.  That the regularizations $V_{m,B}(x)$
 are important for atoms in magnetic fields goes back at
least to Schiff and Snyder \cite{SS} in (1939) and played an
important role in the AHS study \cite {AHS2} of hydrogen. 
In the case $m = 0$ and $B = 1$, (6) can be rewritten as
\begin{eqnarray}
        V_0(x)  & = & \sqrt{\pi} e^{x^2} [1 - \hbox{erf}(x)]  \\
          & = &  \int_0^\infty  \frac{ e^{-u} }{\sqrt{x^2 + u}} du
                  = 2 e^{x^2}\int_x^\infty  e^{-t^2} dt \nonumber
\end{eqnarray}
A comparison with the usual Coulomb potential is given in Figure \ref{fig:V0}.
      
We now describe the correspondence between a restricted
three-dimensional minimization problem for a Hamiltonian with
 a magnetic field, namely,
\begin{eqnarray}
E_0^{m_1 \ldots m_N}(N,Z,B,\alpha) =  \inf_{\Psi_{m_1 \ldots m_N}} 
    \langle \Psi_{m_1 \ldots m_N}, H(N,Z,B,\alpha) \Psi_{m_1 \ldots m_N} \rangle
\end{eqnarray}
and the one-dimensional minimization problem 
\begin{equation}
  E_{0,x}^{m_1 \ldots m_N}(N,Z,B,\alpha) =  \inf_{\Phi} 
     \langle \Phi, \widehat{H}_x^{m_1 \ldots m_N}(N,Z,B,\alpha) \Phi \rangle
\end{equation}
where
\begin{equation}
  \widehat{H}_x^{m_1 \ldots m_N}(N,Z,B,\alpha) =
    \sum_{j=1}^N \left[ -\frac{d^2}{dx_j^2} -Z V_{m_j}^B(x_j) \right] +
        \alpha \sum_{j< k} W_{m_j,m_k}^B(|x_j - x_k|)
\end{equation}
in which no magnetic field is explicitly present.   Since
\begin{eqnarray} 
  \langle \Psi_{m_1 \ldots m_N}, H(N,Z,B,\alpha) \Psi_{m_1 \ldots m_N} \rangle
         =  \langle \Phi,  
       \widehat{H}_x^{m_1 \ldots m_N}(N,Z,B,\alpha) \Phi \rangle  + NB    
\end{eqnarray}
we can conclude that
\begin{eqnarray} 
 \langle \Psi_{m_1 \ldots m_N}, H(N,Z,B,\alpha) \Psi_{m_1 \ldots m_N} \rangle
   >  E_0^{m_1 \ldots m_{N-1}}(N-1,Z,B,\alpha)   + B   \nonumber \\ 
    \iff  
  \langle \Phi, \widehat{H}_x^{m_1 \ldots m_N}(N,Z,B,\alpha) \Phi \rangle
      >  E_{0,x}^{m_1 \ldots m_{N-1}}(N-1,Z,B,\alpha).
\end{eqnarray}
Thus, we have reduced the problem of determining whether or not the 
somewhat artificial problem of whether or not an atom in magnetic
field  $B$ whose electrons have prescribed angular momentum corresponding
to $m_1 \ldots m_N$ has a bound state to that of whether or not a
related one-dimensional system has a bound state. 
Although, we do not consider spin-coupling explicitly, it does follow
from (3) that we can also conclude that
\begin{eqnarray}\label{varequivspin} 
 \langle {\bf \Psi}_{m_1 \ldots m_N}, {\bf H}(N,Z,B,\alpha) 
         {\bf \Psi}_{m_1 \ldots m_N} \rangle
   >  E_0^{m_1 \ldots m_{N-1}}(N-1,Z,B,\alpha)   \nonumber \\ 
    \iff  
  \langle \Phi, \widehat{H}_x^{m_1 \ldots m_N}(N,Z,B,\alpha) \Phi \rangle
      >  E_{0,x}^{m_1 \ldots m_{N-1}}(N-1,Z,B,\alpha).
\end{eqnarray}
where ${\bf \Psi}_{m_1 \ldots m_N}$ is chosen so that all components in
${\bf C}^2$ correspond to spin ``down". 
(This observation, which is a fortuitous consequence of the physical fact
that the coefficient of ${\bf \sigma} \cdot {\bf B}$ is exactly $1$, was
pointed out to us by J.P. Solovej.)

We now consider the regularization of the interaction in the special
case in which $m_j = m_k = 0$.  Recall that 
$\gamma_0^B(y,z) = B \pi^{-1} e^{-B(y^2 + z^2)/2} = B \pi^{-1} e^{-Br^2/2}.$
\begin{eqnarray}
  W_{0,0}^B(|x_1 - x_2|) & = & \int\!\int  dy_1 dz_1\int\!\int  dy_2 dz_2 
    \frac{|\gamma_0^B(y_1,z_1)|^2|\gamma_0^B(y_2,z_2)|^2}
         {\sqrt{(x_1 - x_2)^2 + |(y_1,z_1) - (y_2,z_2)|^2} } \nonumber \\
   & = &   B \int_o^{\infty} dt  
        \frac{2t e^{-Bt^2}}{\sqrt{(x_1 - x_2)^2 + 2t^2} }  \\
   & = & \frac{1}{\sqrt{2}}  V_0^B\left(\frac{|x_1 - x_2|}{\sqrt{2}}\right)
       \nonumber
\end{eqnarray}
where we have made the change of variables 
\begin{eqnarray*}
{\bf s} = \frac{1}{\sqrt{2}} [(y_1,z_1) + (y_2,z_2)] ,~~~ 
      {\bf t} = \frac{1}{\sqrt{2}} [(y_1,z_1) - (y_2,z_2)]
\end{eqnarray*} 
with $s = |{\bf s}|,~ t = |{\bf t}|$, and used the fact that
\begin{eqnarray*}
           |\gamma_0^B(y_1,z_1)|^2 |\gamma_0^B(y_2,z_2)|^2 
& = & B^2\pi^{-2} e^{-B(r_1^2 + r_2^2)} \\
  & = &   B^2\pi^{-2}e^{-B(s^2 + t^2)}   
= |\gamma_0^B({\bf s})|^2  |\gamma_0^B({\bf t})|^2.
\end{eqnarray*}
Thus, the exceedingly simple relation  
$W_{0,0}^B(|x_1 - x_2|) = 2^{-1/2}V_0^B(2^{-1/2}|x_1~-~x_2|)$
follows from the invariance of 
    $|\gamma_0^B({\bf s})|^2  |\gamma_0^B({\bf t})|^2$
under the transformation of
   ${\bf s}, {\bf t}$ to $({\bf s} \pm {\bf t})/\sqrt{2}$.
This unusual invariance, corresponding to the mixing of coordinates
of two particles, will not hold if $m \neq 0$.
Symmetrizing the product or replacing 
$\Psi_{m_1 \ldots m_N}$ by an arbitrary
element of the projection onto the lowest Landau level, 
would require consideration of exchange terms as well.
Therefore, we will only study models corresponding to constraining
all electrons to have angular momentum zero.  Since $\Psi_{0 \ldots  0}$
is then symmetric with respect to exchange of $(y_j,z_j)$ with   
$(y_k,z_k)$, it will have the same permutational symmetry as
$\Phi(x_1 \ldots x_N)$.

Despite the severity of the restriction to $m = 0$, our model seems
well-suited to study of the bounds on the negative ionization.
Integration by parts of (\ref{vmdef}) easily yields 
\begin{eqnarray}
  V_{m+1}^B(x) \leq  V_m^B(x) \leq  V_0^B(x) \leq   1/|x|
\end{eqnarray}
with the difference greatest at the origin and all potentials
satisfying  $V_m^B(x) \approx  1/|x|$ for large $x$.  Moreover,
for any choice of $m_j, ~ m_k$ whenever $|x_j - x_j|$ is large,
 $W_{m_j,m_k}^B(|x_j - x_j|) \approx  1/|x_j - x_j|$ as well.
Thus it
appears unlikely that placing some electrons in Landau levels
with $m \neq 0$ will allow binding if $m = 0$ does not.

With this heuristic background, we study models for N-electron atoms
in one-dimension corresponding to Hamiltonians of the form
\begin{eqnarray}\label{eq:Ham1dim}
  h(N,Z,M) & = &  M \widehat{H}_x^{0 \ldots 0}(N,Z,M^{-2},1) \\
    & = & \sum_{j=1}^N \left[ - \frac{1}{M} \frac{d^2}{dx^2} -Z V_0(x_j) \right] +
         \sum_{j< k} \frac{1}{\sqrt{2}} 
              V_0\left(\frac{|x_j - x_k|}{\sqrt{2}}\right) \nonumber   
\end{eqnarray}
with the ``mass'' $M$ proportional to $B^{-1/2}$.  Because of the
scaling relation $H(N,Z,B,1) = B H(N,ZB^{-1/2},1,B^{-1/2})$,
the only role of the field strength in the one-dimensional
situation, is to reduce the mass by a factor of $B^{-1/2}$.  Thus,
for simplicity, we have set both $B = 1$ and the coupling constant
$\alpha = 1$ leaving the field strength implicit in the mass $M$.

 As observed in \cite{AHS2}, the regularized potentials satisfy
\begin{eqnarray}
  \frac{1}{\sqrt{(m+1)B} + |x|} \leq \frac{1}{\sqrt{(m+1)B + |x|^2}}
      \leq  V_m^B(x) \leq  \frac{1}{|x|}.
\end{eqnarray}
The cut-off potential $V_{\hbox{\scriptsize{cut}}}(x) = 1/(|x| + 1)$ is also of some
interest.  Haines and Roberts \cite{HR} have given explicit solutions to the
eigenvalue problem for the hydrogenic Hamiltonian
$ - \Delta - V_{\hbox{\scriptsize{cut}}}(x)$.  By the above remark,
$V_{\hbox{\scriptsize{cut}}}(x) \leq V_0^1(x) \leq  1/|x|$.

In sections 2 and 3, we study
Hamiltonians of the form (\ref{eq:Ham1dim}) 
in the two cases $V = V_{\hbox{\scriptsize{cut}}}$ 
and $V = V_0$.  For both models we show that the existence of a
bound state implies  $N < 2Z+1+c \sqrt{B}$.

\section {Two-electron systems}

We now discuss in more detail the behavior of the one-dimensional
Hamiltonian (\ref{eq:Ham1dim}) when the number of electrons is
$N = 2$.  Although our discussion is descriptive and non-rigorous,
we believe the insights are useful.  The two-electron Hamiltonian
can be written in the form.
\begin{eqnarray}\label{eq:Ham2elec}
  h(2,Z,B^{-1/2}) = - \sqrt{B} \left[ \frac{d^2}{dx_1^2} +
  \frac{d^2}{dx_2^2} \right] + W(x_1,x_2)
\end{eqnarray}
where
\begin{eqnarray} \label{eq:Wxypot}
 W(x_1,x_2) = -Z V_0(x_1)  -Z V_0(x_2)
      + \frac{1}{\sqrt{2}}  V_0\left(\frac{|x_1 - x_2|}{\sqrt{2}}\right)
\end{eqnarray}

Consider a classical system of three particles on a line with charges
$-1$, $+Z$, $-1$ interacting with the usual inverse square Coulomb 
force (or, equivalently, $q_i q_j/|x_i- x_j|$ potential).
The only potentially stable configuration is one in which
the particles are arranged symmetrically with
the two negatively charged
particles on opposite sides of, and the same distance from, the positive
"nucleus" as shown in Fig. 1.  Even in this case,
the system is stable only for $Z = 1/4$.  For
$Z < 1/4$, both ``electrons" will move off to infinity, while for
$Z > 1/4$, the ``electrons'' collapse into the center.  This suggests
that for $Z > 1/4$ one might be able to show that the system binds
by using a trial function $\Psi_a$ in which the two electrons are localized 
on opposite sides of the nucleus at a distance $a$ sufficiently far
from the center that $V_0(\pm a) \approx 1/a$.  (Permutational
symmetry is not relevant; if such a trial function binds, then a
bosonic system will also bind.)   This picture would require
that $\Delta x_i < a$ for each electron so that, by the uncertainty
principle, the kinetic energy satisfies 
$\langle \Psi_a [- \frac{d^2}{dx_1^2}] \Psi_a \rangle > 1/a^2$.
Thus we estimate
\begin{eqnarray}
   \langle \Psi_a h(2,Z,B^{-1/2}) \Psi_a \rangle  >
      2 \frac {\sqrt{B}}{a^2}  - \frac{2Z}{a} + \frac{1}{2a}.
\end{eqnarray}
Minimizing over $a$ yields $a_{\min} = 2 \sqrt{B}/(Z - \frac{1}{4})$
and
\begin{eqnarray}
   \langle \Psi_a h(2,Z,B^{-1/2}) \Psi_a \rangle >
      - \frac {(Z - \frac{1}{4})^2}{2 \sqrt{B}}.
\end{eqnarray}
The continuous spectrum begins at the ground state energy of
 the corresponding one-electron Hamiltonian 
$h(1,Z,B^{-1/2}) = - \sqrt{B} \frac{d^2}{dx^2} -Z V_0(x)$
which Avron, Herbst and Simon \cite{AHS2} showed is given asymptotically
(for large $B$) by   $E_0(1,Z,B^{-1/2}) =
 - \frac{Z^2}{\sqrt{B}} \left( \log \frac{Z^2}{\sqrt{B}} \right)^2$
Thus, binding would require
\begin{eqnarray}
 (Z - \frac{1}{4})^2 > 2 Z^2  \left( \log \frac{Z}{\sqrt{B}} \right)^2.
\end{eqnarray}
This is obviously false for large $B$, which suggests that the
uncertainty principle prevents the corresponding one-dimensional
system from binding, even when the classical system collapses.
Because $M \sim B^{-1/2}$, large field strength corresponds to small
mass, i.e., as the field strength increases, the electrons become 
``lighter" and more difficult to localize.  Thus we must seek a
different mechanism to explain binding of a one-dimensional
two-electron atom.

Fortunately, our model provides an alternative mechanism for binding.
The  regularization of the potential
at the origin combined with the effective reduction of the interaction
by $1/\sqrt{2}$, permits both electrons to be close to the 
nucleus for $Z > 1/2^{3/2}$ in the sense that 
$-2ZV_0(0) + 2^{-1/2}V_0 < 0$.  Of course, the uncertainty
principle also precludes binding with a trial function in which
both electrons are exactly at the center.  Nevertheless,
we believe that the mechanism for binding is that
the effective reduction in the interaction permits the
two electrons to overlap strongly near the nucleus.

\begin{figure}
\vspace*{7cm}
\hspace*{1.5cm}\psfig{figure=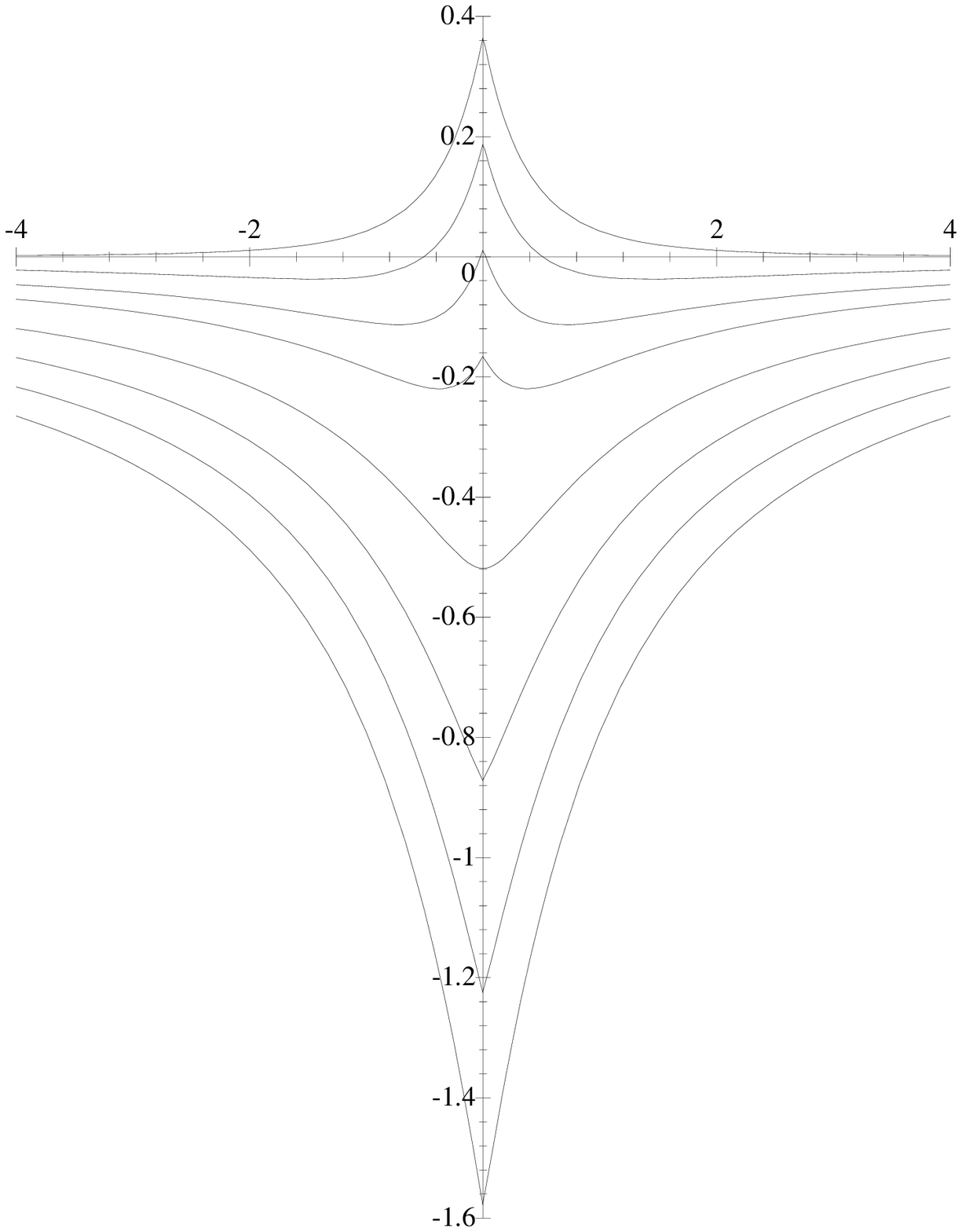,height=1in}
\hspace*{2.5cm}\psfig{figure=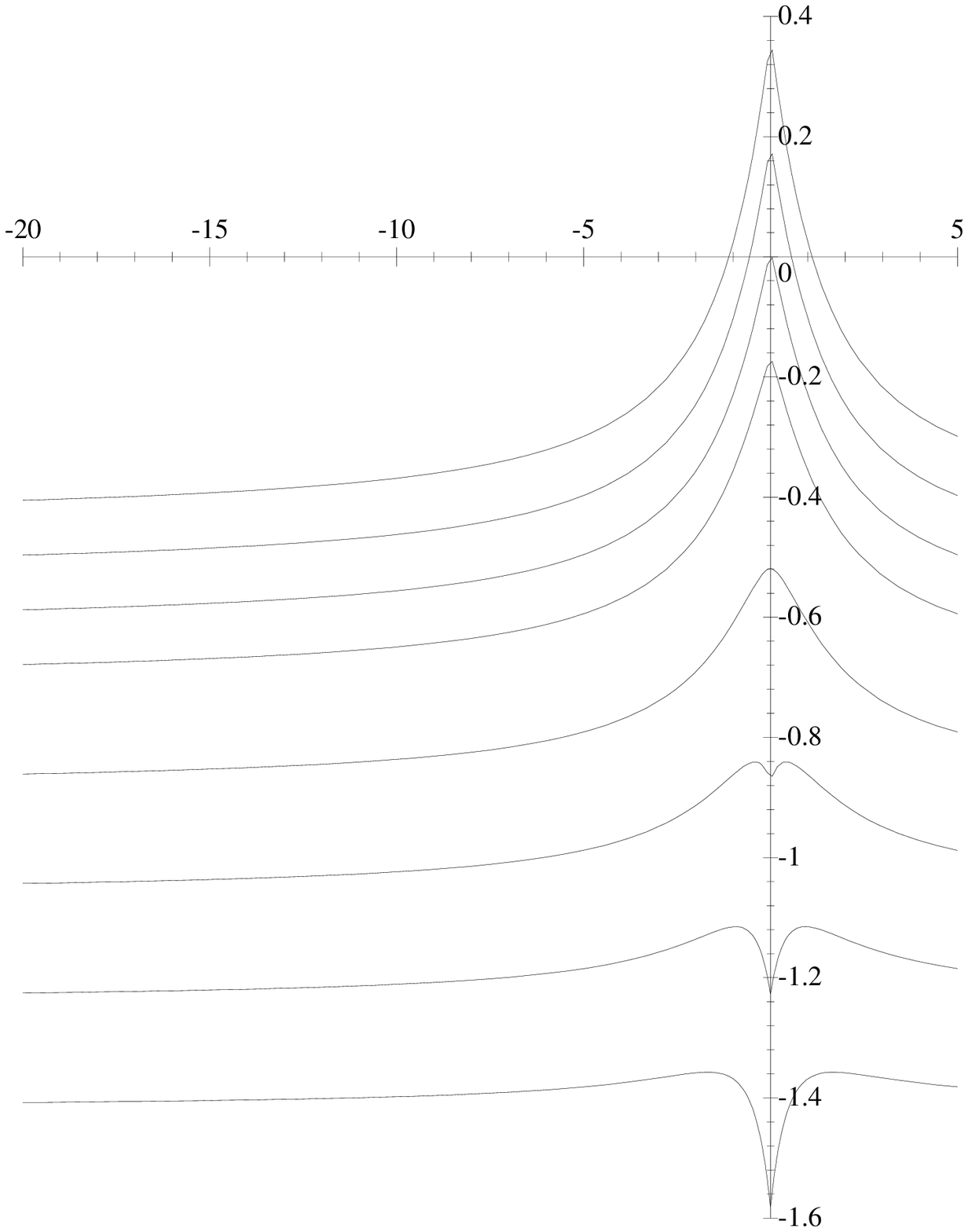,height=1in}
\vskip-3cm
\caption {$W(x,-x)$ and $W(x,0)$ for (reading from the top 
down)  $Z~=~0.25, 0.3, 0.35, 0.4, 0.5, 0.6, 0.7, 0.8$.   
The left  graph, $W(x,-x)$, describes the potential when the electrons 
are on opposite sides of the nucleus with each a distance  
$x$ away. The right graph, $W(x,0)$, describes the potential 
when one electron is fixed at the origin; the left asymptote 
then corresponds to the other electron moving to infinity.}
\label{fig:Wxylines} \end{figure}

The Hamiltonian (\ref{eq:Ham2elec}) can be regarded as describing
either two particles in one-dimension or one particle in the
field of the two-dimensional potential $W(x,y)$ given by 
(\ref{eq:Wxypot}).  Regardless of our viewpoint, the continuous
spectrum will begin at the AHS \cite{AHS2} estimate of 
$ - \frac{Z^2}{\sqrt{B}} \left( \log \frac{Z^2}{\sqrt{B}} \right)^2$.
Since this is close to zero for large $B$, we will regard
$W(x,y)$ as attractive where it is negative and repulsive
where it is positive.
One might expect $W(x,y)$ to have its minimum on the line $y = -x$,
corresponding to electrons on opposite sides of the nucleus.  The
actual situation, which is described below, is more complex.
Plots of $W(x,-x)$ and $W(x,0)$ are shown in Fig. 2 for 
$Z = 0.25, 0.3, 0.35 \approx \frac{1}{2\sqrt{2}}, 0.4, 0.5,
 0.6, 0.7 \approx  \frac{1}{\sqrt{2}}, 0.8.$   The graph of
$W(x,0)$ is deliberately asymmetric so that the left asymptote,
$W(-\infty,0)$ can be compared to the minimum of $W(x,-x)$
at $x = 0.$  We now discuss this behavior in more detail,
noting that all the qualitative
features can also be verified analytically.
\begin{itemize}
\item[$\bullet$] $\frac{1}{4} < Z < \frac{1}{2\sqrt{2}} : $   
The potential does have a pair of weak minima along the line $y = -x$;
however, closer examination of the full two-dimensional 
potential shows that these are not true minima, but saddle
points for $W(x,y)$.  The potential is repulsive at the origin
and only weakly attractive elsewhere.
\item[$\bullet$] $\frac{1}{2\sqrt{2}} < Z < \frac{1}{2} : $ The potential is now
attractive at the origin.  However, as above, the two weak
minima on the line $y = -x$ correspond to saddle points of $W(x,y)$.
\item [$\bullet$] $\frac{1}{2} < Z  < \frac{1}{\sqrt{2}} : $ The two weak   
minima on the line $y = -x$ have now coalesced into a true
(two-dimensional) minimum at the origin.  However, this is only a local minimum.
$W(0,0) > W(0,\infty)$ so that the energy will decrease if one of 
the particles remains at the origin while the other goes off to infinity.
Fig. 2 suggests that the minimum along the $x=0$ and $y=0$ lines
is too shallow to prevent one of the electrons from tunnelling
through to infinity.  Thus, we expect resonances, but not bound
states in this region.
\item[$\bullet$] $ Z > \frac{1}{\sqrt{2}} : $  The potential has a true
minimum at the origin.   
\end{itemize}

\begin{figure}
\vskip3.5cm
\hskip2in\psfig{figure=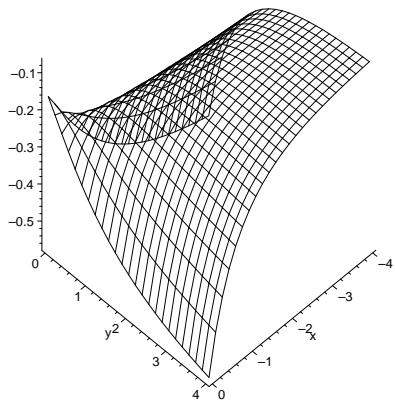,height=1in}
\vskip-0.5cm
\caption{Saddle point of $W(x,y)$ for $Z=0.4$ in the quadrant $x < 0$, $y > 0.$
 \label{fig:saddle}}
\end{figure}

Thus, the behavior of the potential $W(x,y)$ strongly suggests
that, at least for sufficiently large $B$,
binding occurs for $Z > \frac{1}{\sqrt{2}} \approx 0.7$ and that 
the mechanism which permits this is the reduction in strength
of the repulsion by $1/\sqrt{2}$ which permits both electrons to
simultaneously sit near the nucleus.  The behavior of the
regularized potential model for a two-electron atom in one dimension
seems to be very different from that of classical Coulomb particles
confined to a line.  The binding cut-off of $Z_c \approx 0.7$ 
represents a very modest ``binding enhancement'' when compared
to the value of $Z_c \approx 0.9$ obtained in \cite{BFHM}
for a two-electron atom in three dimension in the absence of a
magnetic field. [An ``enhancement'' to the level of
$2 = N  > 2Z+1$ would require $Z_c < 0.5$.]
\bigskip

\section {Kinetic Energy and Lieb's Strategy}

We now describe the elegant strategy used by Lieb \cite{Lb} 
(see also section 3.8 of \cite{CFKS}) to show that for
real atoms the maximum number of electrons that can be bound is
less than $2Z+1$ irrespective of permutational symmetry.  The essence of
Lieb's argument is to show that one can replace the variation over
N-electron wave functions $\Psi({\bf r}_1, \ldots {\bf r}_N)$ by
a variation over density matrices of the form  
\begin{eqnarray}
\lefteqn{\Gamma({\bf r}_1, \ldots {\bf r}_N ; {\bf s}_1, \ldots {\bf s}_N)} \\
~~ & = & \sum_{j=1}^N 
    \overline{[\nu({\bf r}_j)]^{1/2} \Psi({\bf r}_1, \ldots {\bf r}_N)} 
[\nu({\bf s}_j)]^{1/2} \Psi({\bf s}_1, \ldots {\bf s}_N) 
   \nonumber
\end{eqnarray}
where $\nu({\bf r}) > 0$ is a strictly positive function
which will be chosen later. 
Note that, even though the functions $\Psi \nu_j^{1/2}$ 
are not orthonormal,
  $\T \Gamma A = \sum_j 
     \langle [\nu({\bf r}_j)]^{1/2} \Psi, A [\nu({\bf r}_j)]^{1/2} \Psi \rangle$
for any operator $A$.
Despite the introduction of the function $\nu$,
which will be chosen later, the $\Psi$  used in
$\Gamma$ satisfies the same symmetry or domain
constraints as the original minimization problem.

It will be useful to describe Lieb's argument in the rather
general situation of an N-particle Hamiltonian, $H_N$ with
the structure $H_N = H_{N-1}^j  + K_j + V_{\textstyle{\hbox{\scriptsize{int}}}}^j$ where
$K_j = |{\bf P}_j + {\bf A}|^2$ denotes the kinetic energy,
$H_{N-1}^j$ is the $N-1$-electron Hamiltonian in which the j-th
electron is omitted and $V_{\textstyle{\hbox{\scriptsize{int}}}}^j$ is a potential which
describes the interaction of the j-th electron with the rest
of the system.  Lieb studied $H_N = H(N,Z,0,1)$ while
we will consider $H_N =  h(N,Z,M)$.  In our case,
\begin{eqnarray}
V_{\textstyle{\hbox{\scriptsize{int}}}}^j = 
       -ZV(x_j) + \sum_{k \neq j} 2^{-1/2} V(2^{-1/2}|x_j - x_k|)
\end{eqnarray}
 
Now assume that $H_N$ actually has a bound state so that
there is a $\Psi_0$ satisfying $H_N \Psi_0 = E_0 \Psi_0$.  This
implies that for {\em any} function $\nu$ and {\em any} choice of $j$, 
\begin{eqnarray} 
    \nu({\bf r}_j) H_N \Psi_0({\bf r}_1, \ldots {\bf r}_N) 
    =  E_0 \nu({\bf r}_j) \Psi_0({\bf r}_1, \ldots {\bf r}_N) 
\end{eqnarray}
as well.  Hence
\begin{eqnarray}\label{eq:gamma.en} 
     \langle \Psi_0,  \nu({\bf r}_j) H_N \Psi_0 \rangle = 
   E_0   \langle \Psi_0,   \nu({\bf r}_j) \Psi_0\ \rangle .
\end{eqnarray}
Then if  $\Gamma_0$ is the density matrix of the above form corresponding
to the ground state $\Psi_0$, and $\nu_j$ denotes $\nu({\bf r}_j)$
\begin{eqnarray}\label{eq:gammavarprin}
\lefteqn  {E_0(N) \T \Gamma_0   =   
           E_0(N) \sum_{j=1}^N \langle  \Psi \nu_j\Psi \rangle} ~~~ \nonumber \\
 & = & \T \Gamma_0 H_N + \sum_{j=1}^N \left[ \langle\Psi \nu_j K_j \Psi \rangle 
          - \langle  \nu_j^{1/2} \Psi K_j \nu_j^{1/2}\Psi \rangle \right] 
     \nonumber  \\
 & = & \sum_j \langle  \nu_j^{1/2} \Psi,   \left [ H_{N-1}^j 
       + V_{\textstyle{\hbox{\scriptsize{int}}}}^j \right]   \nu_j^{1/2} \Psi \rangle  +
        \sum_{j=1}^N  \langle\Psi \nu_j K_j \Psi \rangle  \label{eq:Lbvar} \\
  & \geq &  E_0(N-1) \T \Gamma_0 +   \sum_j \left[
    \langle \nu_j^{1/2}\Psi, V_{\textstyle{\hbox{\scriptsize{int}}}}^j ~\nu_j^{1/2}\Psi \rangle
   + \langle\Psi \nu_j K_j \Psi \rangle \right]    \nonumber
\end{eqnarray}
The inequality in (\ref{eq:Lbvar}) above follows from the variational
principle for $H_{N-1}^j$.  Although  $\langle\Psi \nu_j K_j \Psi \rangle $
need not be real in general, a careful analysis of the argument
above shows that when $\Psi$ is an eigenstate of $H_N$ it
is real because it can be written as the difference of two real
quantities.  In Lieb's original formulation, $\nu$ was chosen
so that this real quantity satisfied
$\langle\Psi \nu_j K_j \Psi \rangle > 0.$  Then if 
$ \sum_j \langle \nu_j^{1/2}\Psi V_{\textstyle{\hbox{\scriptsize{int}}}}^j \nu_j^{1/2}\Psi \rangle > 0$
it follows that  $ E_0(N) \T \Gamma_0 >  E_0(N-1) \T \Gamma_0$
which contradicts the requirement for binding of
$ E_0(N) < E_0(N-1) $.  This reduces the problem of showing
that $H(N,Z,0,\alpha)$ has no bound states in the absence of
magnetic fields to that of showing that  
$ \sum_j \langle \nu_j^{1/2}\Psi V_{\textstyle{\hbox{\scriptsize{int}}}}^j \nu_j^{1/2}\Psi \rangle > 0.$
However, in a constant magnetic field the continuum begins
at $B + E_0(N-1)$ so that a similar reduction would require
the stronger condition  $\langle\Psi \nu_j K_j \Psi \rangle > B$.  

In order to study  $\langle\Psi \nu_j K_j \Psi \rangle$ in more
detail in different situations, it is convenient to
write $\nu_j$ as $g^2$ and observe that formally
\begin{eqnarray}\label{eq:kegena} 
\lefteqn{\langle \Psi g^2, |i\nabla_j + {\bf A}|^2 \Psi \rangle  
   = \langle [i\nabla_j + {\bf A}](g\Psi) g , 
          [i\nabla_j + {\bf A}](g \Psi) g^{-1} \rangle} \nonumber \\
  & = & 
  \langle [i\nabla_j + {\bf A}]g\Psi, [i\nabla_j + {\bf A}] (g\Psi) \rangle
       - \int |g\Psi|^2 \left|\frac{ \nabla g}{g}\right|^2    \\
    & & \pm ~2i~\Re \langle \Psi \nabla g,   [i\nabla_j + {\bf A}] (g\Psi) .
  \rangle   \nonumber
\end{eqnarray}
Note that the localization error,
$- \int |g\Psi|^2 \left|\frac{ \nabla g}{g}\right|^2 =   
\int |g\Psi|^2 \nabla g \cdot \nabla (g^{-1})$,
shows a certain symmetry between $g$ and $g^{-1}.$
If we now let  $g = \nu^{1/2}$, (\ref{eq:kegena}) implies
\begin{eqnarray}\label{eq:kegenb}
  \Re \langle\Psi \nu_j K_j \Psi \rangle =   \langle \Psi \nu_j^{1/2},
 \left[ K_j - \left| \frac{\nabla_j \nu_j}{2\nu_j} \right|^2 \right]
           \Psi \nu_j^{1/2} \rangle .
\end{eqnarray}
For the typical choice $\nu = [V({\bf r})]^{-1}$, one has
$\nabla \nu({\bf r})/\nu({\bf r}) = - \nabla V({\bf r})/V({\bf r})$
which yields
\begin{eqnarray}\label{eq:kegenc}
  \Re \langle\Psi \nu_j K_j \Psi \rangle =   \langle \Psi \nu_j^{1/2},
   \left[ K_j - \left| \frac{\nabla_j V_j}{2V_j} \right|^2 \right]
           \Psi \nu_j^{1/2} \rangle 
\end{eqnarray}
in accordance with the symmetry between $g$ and $g^{-1}$ noted above.
Because, as discussed above, we will only be concerned with
applications for which $\langle\Psi \nu_j K_j \Psi \rangle$ is real 
we will henceforth omit the $\Re$.  We now  discuss
several cases in more detail using ${\bf d}$ to denote the dimension
of the space on which $K_j$ acts.
(The original proof of Lieb used an argument which originated with   
Benguria (see Lemma 7.20 of \cite{LbRMP}) in the spherically symmetric   
case to show directly that   
$\Re \langle \nu \Psi (-\Delta) \Psi \rangle > 0$.  The variant
given here is due to Hoffman-Ostenhof \cite{HO2}.  For other strategies see
\cite{Baum,CFKS,Ich})
  
\noindent $\bullet~ {\bf A = 0, d \geq 3: }$ In this case we are
interested in  $\nu = [V({\bf r})]^{-1} $ for potentials, (particularly the
usual Coulomb potential $V({\bf r}) = 1/|{\bf r}|$) which
satisfy $\left| \frac{\nabla V({\bf r})}{V({\bf r})} \right| 
     \leq \frac{1}{|{\bf r}|}.$ Then $K = -\Delta$ and
(\ref{eq:kegenc}) becomes
\begin{eqnarray}\label{eq:ke3dim}
  \langle \Psi \nu_j K_j \Psi \rangle &=&
        \langle \Psi \nu_j (-\Delta_j) \Psi \rangle \nonumber \\
  & \geq &  \langle \Psi \nu_j^{1/2},  
        \left[ -\Delta_j - \frac{1}{4 |{\bf r}|^2}
              \right] \Psi \nu_j^{1/2} \rangle \geq 0
\end{eqnarray}

\noindent $\bullet{\bf A \neq 0, d = 3: }$  In this case we can only
conclude from (\ref{eq:kegenb})that
\begin{eqnarray}
  \langle\Psi \nu_j K_j \Psi \rangle & \geq &  B \|\Psi \nu_j^{1/2}\|^2
   -  \langle \Psi \nu_j^{1/2},  \left[ \frac{\partial^2}{\partial x_j^2} - 
     \left| \frac{\nabla_j \nu_j}{2\nu_j} \right|^2 \right] 
         \Psi \nu_j^{1/2} \rangle \\ 
   & \geq &  B \|\Psi \nu_j^{1/2}\|^2
  - \langle \Psi \frac{|\nabla_j \nu_j|^2}{4\nu_j}  \Psi  \rangle. \nonumber
\end{eqnarray}
We could instead have proceeded as in (\ref{eq:ke3dim}) above to obtain, 
$ \langle\Psi \nu_j K_j \Psi \rangle > 0$, but we cannot
conclude that 
$ \langle\Psi \nu_j K_j \Psi \rangle  \geq   B \|\Psi \nu_j^{1/2}\|^2.$
The kinetic energy
 $P_x^2 = \frac{\partial^2}{\partial x_j^2}$ in the field direction is 
not able to control a 3-dimensional potential, such as $1/{\bf r}^2$
arising from $\left| \frac{\nabla_j \nu_j}{2\nu_j} \right|^2 $.
Our one-dimensional models circumvents this problem
because the choice of product state involving Landau functions
ensured that the $P_y^2 + P_z^2$ terms took care of the $B$, leaving
$P_x^2$ to deal with a one-dimensional potential$[V'(x)/2V(x)]^2$ 
as described below.
   
\noindent$\bullet~$ {\bf A = 0, d=1:} In one dimension, ${\bf A}$ plays   
no role.  Although we will still choose
$\nu = [V(|r|)]^{-1} $ with potentials which satisfy
$\left|\frac{V^\prime(x)}{x}\right|^2 < \frac{1}{|x|^2}$ it is not true in
one or two-dimensions that $-\Delta - |2x|^{-2} > 0$.  Instead
we will treat 
$- \left| \frac{V^{\prime}}{2V} \right|^2$ as a potential in  
\begin{eqnarray}
 \langle \Psi \nu_j (-\Delta_j) \Psi \rangle =
    \langle \Psi \nu_j^{1/2},  \left[ -\frac{d^2}{dx_j^2} -  
    \left| \frac{V^{\prime}(x_j)}{2V(x_j)} \right|^2 \right] 
              \Psi \nu_j^{1/2} \rangle .
\end{eqnarray}
Now in one-dimension the everywhere negative potential 
$- \left| \frac{V^{\prime}}{2V} \right|^2 $ will always give rise to
a bound state of 
$ -\frac{d^2}{dx^2} - \left| \frac{V^{\prime}}{2V} \right|^2$.  
Thus we can only conclude
that  $\langle \Psi \nu_j (-\Delta_j) \Psi \rangle \geq
   -\epsilon \| \Psi \nu_j^{1/2} \|^2 $ where $ -\epsilon$ is the lowest
eigenvalue of this operator.  However, the bound 
\begin{eqnarray}
  \langle \Psi \nu_j (-\Delta_j) \Psi \rangle \geq
     - \omega \| \Psi \|^2
\end{eqnarray}
where $\omega = \sup_x | \nu^\prime(x)|^2 /4\nu(x)$
will be more useful.   Notice that, unlike the 3-dimensional case
where the lower bound is zero, when applying this result to
$H_N = h(N,Z,M)$ we will need
to take into account the fact the entire kinetic energy term
is multiplied by $1/M$ (or $\sqrt{B}$).

\section {Bound on the Maximum Negative Ionization}

We now apply Lieb's strategy, which yields $N_{\max}(Z,0) < 2Z+1$
for atoms without magnetic fields, to our one-dimensional models
for systems in strong magnetic fields.  We obtain the following
result.
\begin{thm}  The maximum number of electrons $N_{\max}$ for which
a Hamiltonian $h(N,Z,B^{-1/2})$ of the form (\ref{eq:Ham1dim}) has
a bound state satisfies
 $N_{\max} < 2Z+1 + c \sqrt{B}$ for some constant $c$.
\end{thm}
In the interesting case $B = O(Z^3)$ (which is the boundary of the   
LSY hyperstrong limit region in \cite{LSY2}), this yields a bound of the   
form $N_{\max} < 2Z+ c Z^{3/2}$, rather than  a linear
one of the form  $N_{\max} < cZ$ or the expected optimal
$N_{\max} < 2Z + o(Z)$.

We apply the strategy of Section 3 with $H_N = h(N,Z,M)$ and $\nu = 1/V$.
In this case, the analysis of expectations of 
$V_{\textstyle{\hbox{\scriptsize{int}}}}^j$
is straightforward because 
$ \langle \nu_j^{1/2} \Psi, V   \nu_j^{1/2} \Psi \rangle  =
       \langle \Psi, \Psi \rangle$ .
\begin{eqnarray}
\lefteqn{ \sum_j \langle  \nu_j^{1/2}\Psi
        V_{\textstyle{\hbox{\scriptsize{int}}}}^j \nu_j^{1/2}\Psi \rangle}~~~
     \nonumber \\  
 & = &   -Z \sum_j \| \Psi \|^2  + \sum_j \sum_{k \neq j}
      \langle \Psi, 2^{-1/2} V(2^{-1/2} |x_j - x_k|) V(x_j)^{-1} \Psi \rangle \\
  & = & -NZ \| \Psi \|^2 + \sum_{j < k} \langle \Psi, 2^{-1/2}     
      V(2^{-1/2} |x_j - x_k|)[ V(x_j)^{-1} + V(x_k)^{-1}] \Psi \rangle. \nonumber
\end{eqnarray}
Thus, {\em if} the potential satisfies 
\begin{eqnarray}\label{triineqsubs}
   2^{-1/2}V(2^{-1/2} |x_j - x_k|)[ V(x_j)^{-1} + V(x_k)^{-1}] > 1,
\end{eqnarray}
then
\begin{eqnarray}
 \sum_j \langle  \nu_j^{1/2} \Psi 
        V_{\textstyle{\hbox{\scriptsize{int}}}}^j \nu_j^{1/2} \Psi \rangle \geq
  [- NZ + N(N-1)/2 ] \| \Psi \|^2.
\end{eqnarray}

For the cut-off potential $V(x) = 1/(|x|+1)$, (\ref{triineqsubs})
follows easily from the triangle inequality since
\begin{eqnarray*} 
2^{1/2}[V(2^{-1/2} |w-x|)]^{-1} &=& |w-x| + 2^{1/2}  \\
   & \leq & |w| + 2^{1/2}  + |x| + 2^{1/2}  = [ V(w)]^{-1} + [V(x)^{-1}]
\end{eqnarray*}
For the regularized potential $V_0(x)$ we will use instead
the  convexity of $\nu_0(x) = [V_0(x)]^{-1}$.  (Because the proof 
\cite{W,SzW}
of this essential fact, although elementary, is rather delicate and not
readily accessible, we provide a sketch in the Appendix.)
Thus we find,
\begin{eqnarray}
    [ V_0(w)]^{-1} + [V_0(x)^{-1}] & \geq &  2 [V_0(|w-x|/2)]^{-1} \nonumber \\
      & \geq &  2^{1/2} [V_0(2^{-1/2}|w-x|]^{-1}
\end{eqnarray}
where the first inequality used $V_0(x) = V_0(|x|)$ as well as the
convexity of  $\nu_0(x) = [V_0(x)]^{-1}$ and  the second inequality follows
from
\begin{eqnarray*}
  V_m(2^{-1/2}y) 
   &=&  [m!]^{-1} \int_0^\infty \frac{t^m e^{-t}}{\sqrt{y^2/2 + t}}  ~dt \\
  & = & 2^{1/2} [m!]^{-1}\int_0^\infty  \frac{t^m e^{-t}}{\sqrt{y^2 + 2t}}  ~dt \\
  & \leq & 2^{1/2} V_m(y)  
\end{eqnarray*}
with  $y = 2^{-1/2} |w-x|$ and $m=0$.

Thus we can conclude that, for both the cut-off potential
$V_{\hbox{\scriptsize{cut}}}(x) = 1/(|x| + 1)$ 
and the regularized potential $V_0(x)$,
the one dimensional Hamiltonian $h(N,Z,M)$ satisfies
\begin{eqnarray}\label{eq:lastcond}
\lefteqn{\T \Gamma_0 h(N,Z,M)}~~~ \\ & \geq & E_0(N,Z,M) \T \Gamma_0 + 
      [- NZ + N(N-1)/2 -N \omega/M ]\| \Psi \|^2. \nonumber
\end{eqnarray}
The second term in (\ref{eq:lastcond}) will be positive if
$(N-1)/2 > Z + \omega/M.$ 
For both potentials,  
$\omega = \sup_x | \nu^\prime(x)|^2 /4\nu(x)$
is given by $\lim_{x \rightarrow 0} |\nu^\prime(x)|^2 /4\nu(x).$ 
This yields, $\omega_{\hbox{\scriptsize {cut}}} = 0.25$  for the cut-off
potential and $\omega_0 =  \pi^{-3/2} < 0.18 $ 
for the regularized potential $V_0(x).$   The latter follows
from  $\omega = \nu(0)^3 =  \pi^{-3/2}$ and the fact that,
as shown in the Appendix,
$|\nu^\prime(x)|^2 /4\nu(x) = \nu(x)[\nu(x)-|x|]^2$ is decreasing
for $x > 0$. 
Thus, in both cases, the second term in (\ref{eq:lastcond}) will 
be positive if $N > 2Z+1+1/2M$.  Since binding implies
 $ \T \Gamma_0 h(N,Z,M) \leq  E_0(N,Z,M) \T \Gamma_0 $
and $M$ is proportional to $B^{-1/2}$,
we have shown that one can find a constant $c$ such that
our model one-dimensional system does not have bound states if  
$N \geq 2Z+1+c\sqrt{B}$ or $N_{\max} < 2Z+1+c\sqrt{B}$.

Although this bound is not optimal,
it should be remembered that we are analyzing a Hamiltonian
with the structure
\begin{eqnarray} 
h(N,Z,B^{-1/2}) = h(N-1,Z,B^{-1/2}) + B^{1/2}K_j + 
    V_{\textstyle{\hbox{\scriptsize{int}}}}(x_j)
\end{eqnarray}
and that the factor $\sqrt{B}$ multiplying the kinetic energy is
something of a two-edged sword.  On the one hand, it raises the
energy of the effective one-electron Hamiltonian
$B^{1/2}K_1 + \int V_{\textstyle{\hbox{\scriptsize{int}}}}
|\psi(x_1,x_2,\ldots,x_N)|^2 dx_2 \ldots dx_N $; 
on the other, it multiplies any error arising from the kinetic
energy -- whether using Lieb's strategy or the Ruskai-Sigal
localization approach \cite{CFKS,Rusk1,Rusk2,Sig} -- by $\sqrt{B}.$
Since such correction terms are typically negative, the result
of such treatments is to perversely magnify the negative error associated
with a positive term otherwise expected to raise the energy.
\bigskip

\noindent{\bf Acknowledgment}  It is a pleasure to thank Professor
Pierre Duclos for many helpful discussions, including a critical
observation about the form of the interaction, and for his
hospitality during visits to the the Universit\'e de Toulon et du Var
and the Centre Physique Th{\'e}orique at Luminy-Marseille. 
We are also grateful to Professor Phillip Solovej for several helpful
comments and to Professor Elisabeth Werner for communicating
the results in \cite{SzW} and the strategy used to prove inequality
(\ref{ineq:g3Vg4}) in the Appendix.  
The second author [MBR] appreciated the hospitality 
at the Universit\'e de Reims and at Georgia Tech where parts of this
work were done.

\appendix\section {Convexity and Properties of $1/V_0(x)$}
\begin{figure}
\vskip3cm
\hskip4cm\psfig{figure=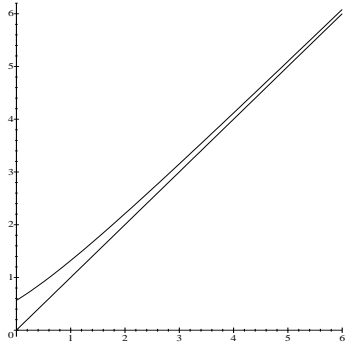,height=1in,width=1in}
\vskip-0.5cm
\caption{Comparison of $1/V_0(x)$ and $x$ for $x \geq 0$.}
\label{fig:convex}
\end{figure}

  Because $ \nu(x) = 1/V_0(x)$ is nearly linear (see Figure \ref{fig:convex}),
the proof of its convexity is somewhat delicate.  It will follow from
the upper bound in the following pair
of inequalities, which are of some interest in their own right.
\begin{eqnarray}\label{ineq:g3nug4}
 \frac{3x + \sqrt{x^2+4}}{4} < \nu(x) <  \frac{2x + \sqrt{x^2+3}}{3}.
\end{eqnarray}
We now restrict attention to $x > 0$ and define
\begin{eqnarray}\label{eq:gkdef}
 g_k(x) = \frac{k}{(k-1)x + \sqrt{x^2+k}}
\end{eqnarray}
so that (\ref{ineq:g3nug4}) is equivalent to
\begin{eqnarray}\label{ineq:g3Vg4}
  g_4(x) > V_0(x) > g_3(x).
\end{eqnarray}
which is an improvement on the classical inequalities of Komatsu
\cite{IM}.  The upper bound and convexity were established independently 
by Wirth \cite{W} and by Szarek and Werner \cite{SzW}; the proof of
lower bound which follows was communicated to the authors by
E. Werner.

  We now observe that $V_0(0) = \sqrt{\pi}$ and $V_0$ satisfies the 
  differential equation
\begin{eqnarray}\label{diffeqV}
  V_0'(x) = 2[x V_0(x) -1]
\end{eqnarray} 
and look for analogous behavior for $g_k.$  
Since  $g_k^{\prime}(x) = -[g_k(x)]^2~
     \frac{x + (k-1)\sqrt{x^2+k}}{k~\sqrt{x^2+k} } $ and
    $ x g_k(x) -1 = \frac{-1}{x + \sqrt{x^2+k}}~g_k(x)$,
one can verify that
\begin{eqnarray}\label{ineq:deriv}
   \lefteqn{ g_k'(x) > 2[x g_k(x)  - 1] } ~~~~~~~~ \nonumber \\
        & \iff &  \frac{k}{\sqrt{x^2+k}}~
    \frac{(k-1)\sqrt{x^2+k} + x}{(k-1)x + \sqrt{x^2+k}} < 
      \frac{2k}{x + \sqrt{x^2+k}} \nonumber \\
           & \iff & (k-2) x^2 + k (k-3) < (k-2) x \sqrt{x^2+k} \nonumber
        \\ & \iff & x^2 (k-2)(k-4) + k(k-3)^2 < 0.
 \end{eqnarray}
 For $k=3$, we use the first equivalence to conclude from   
$x^2 < x \sqrt{x^2+3} $ that $g_3'(x) > 2[xg_3(x)  - 1] $;   
for $k=4$, we use the second to conclude
$k(k-3)^2 > 0 \Rightarrow  g_4'(x) < 2[xg_4(x)  - 1].$

\begin{figure}
\vskip3cm
\hskip5cm\psfig{figure=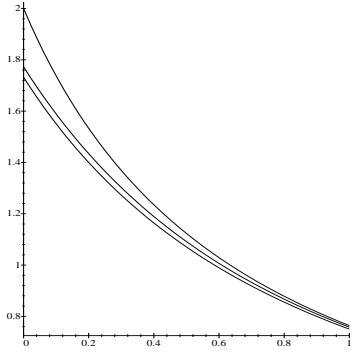,height=1in,width=1in}
\vskip-0.5cm
\caption{Graphs showing $g_4(x) > V_0(x) > g_3(x)$ for $x > 0$.}
\label{fig:comparebd}
\end{figure}

For $k=3$, let $h_3(x) = V_0(x) - g_3(x)$. It is easily to verify that
$h_3(0) = \sqrt{\pi} - \sqrt{3} > 0$ and that   
$\lim _{x \to \infty } |h_3(x)| = 0 $. The first   
equivalence in (\ref{ineq:deriv}) implies
that $x^2 < x \sqrt{x^2+3} \Rightarrow g_3'(x) > 2[xg_3(x)  - 1]$
so that $h'_3(x) < 2x [V_0 (x) - g_3(x)] = 2x h_3(x) ~\forall x.$
Now suppose that for some $x= a$, $h_3(a) < 0$. Then $h'_3(a) < 0$ and
it follows that $h_3(x)$ is decreasing for $x > a$, which 
contradicts $\lim_{x \rightarrow \infty} h_3(x) = 0.$  This
proves the lower bound in (\ref{ineq:g3Vg4}).  The upper bound
is proved similary \cite{SzW} except that one now shows that
$h_4(x) = g_4(x) - V_0(x) > 0.$
It is interesting to note that the upper bound in
(\ref{ineq:g3Vg4})  is optimal, but the lower bound can be
improved \cite{RuW} to $g_{\pi}(x) < V_0(x).$
The tightness of these bounds is evident in Figure \ref{fig:comparebd}.

In order to use this to establish the convexity of $\nu(x)$, 
we note (\ref{diffeqV}) implies that   
 $\nu(x) = [V_0(x)]^{-1}$ satisfies  
\begin{eqnarray}\label{diffeq}   
 \nu'(x) = 2\nu(x)[\nu(x) - x].   
\end{eqnarray}   
Then   
\begin{eqnarray}\label{deriv2}   
 \nu''(x) & =  &4 \nu\nu' -2 \nu'x -2\nu    
     = 4\nu(\nu-x)(2\nu-x) -2\nu \nonumber \\   
  & = & 8\nu [ (\nu - \textstyle{\frac{3}{4}} x)^2 - \textstyle{\frac{1}{16}}
            (x^2+4)]  \\   
   & = & 8\nu \left[ (\nu(x) - 1/g_4(x) + \textstyle{\frac{1}{4}} \sqrt{x^2+4})^2 - 
       (\textstyle{\frac{1}{4}} \sqrt{x^2+4})^2  \right]. \nonumber 
\end{eqnarray}
Since (\ref{ineq:g3nug4}) implies
$\nu(x) > 1/g_4(x)$, we can conclude that $\nu''(x)  > 0.$

To show that
$|\nu^\prime(x)|^2 /4\nu(x) = \nu(x)[\nu(x)-|x|]^2$  is decreasing for 
$x > 0$, we note that (\ref{diffeq}) yields
\begin{eqnarray*}
 \frac{d}{dx} \left[ \nu (\nu-x)^2  \right] & = &
   2 \nu (\nu - x) [3 \nu^2 - 4 \nu x + x^2 -1] \\
     & = & 6 \nu (\nu - x) \left[ (\nu - \frac{2x}{3})^2 - \frac{x^2 + 3}{9} 
     \right]  < 0.
\end{eqnarray*}
and (\ref{ineq:g3nug4}) implies that the expression in square brackets
is negative.

It is useful to have bounds on the extent to which
$\nu(x)$ and $V_0(x)$ deviate from $|x|$ and the Coulomb potential,
respectively, when $x$ is large.  The upper bound in (\ref{ineq:g3nug4})
implies 
$\left|\nu(x) - |x| \right| < g_4(x) - x \leq 1/2|x|$ 
while the lower bound yields 
$\left|V_0(x) - 1/|x| \right| < 1/x - g_3(x) < 1/2x^3.$  
The latter can also be proved directly from the integral
representation (\ref{vmdef}).


\begin{thebibliography}{~~}

\bibitem{AHS1}  Y. Avron, I. Herbst, and B. Simon, 
``Schr\"ošdinger Operators with Magnetic Fields I. 
General Interactions'' {\em Duke Math. J.} {\bf 45} 847-883 (1978).

\bibitem{AHS2} Y. Avron, I. Herbst, and B. Simon, ``Strongly Bound States
of Hydrogen in Intense Magnetic Field'' {\em Phys. Rev. A} {\bf 20}, 
2287-2296 (1979).

\bibitem{AHS3} Y. Avron, I. Herbst, and B. Simon 
``Schr\"ošdinger Operators with Magnetic Fields III. 
Atoms in Homogeneous Magnetic Field'' {\em Commun. Math. Phys.} 
{\bf 79}, 529-572 (1981).

\bibitem{BFHM}	J.D. Baker, D.E. Freund, R.N. Hill and J.D. Morgan III
 ``Radius of Convergence and Analytic Behavior of the  1/Z  ExpansionÓ
{\em Phys. Rev. A} {\bf 41}, 1247-1273 (1990).

\bibitem{Baum} B. Baumgartner, ``On the Degree of Ionization in the
TFW Theory'' {\em Lett. Math. Phys.}  {\bf 7}, 439-441 (1983).

\bibitem{CFKS}
H.L. Cycon, R.G. Froese, W. Kirsch, and B. Simon
{\em Schr\"{o}dinger Operators} (Springer-Verlag, 1987).

\bibitem{FS}
C.L. Fefferman and L.A. Seco, ``Asymptotic Neutrality of Large Ions''
{\em Commun. Math. Phys.} {\bf 128}, 109-130 (1990).

\bibitem{HR} I.K. Haines and D.H. Roberts,  ``One Dimensional
 Hydrogen AtomÓ  {\em Am. J. Phys.}  {\bf 37}, 1145-1154 (1969). 

\bibitem{HO2} T. and M. Hoffman-Ostenhof, private communication.

\bibitem{Ich} T. Ichinose, ``Note on the Kinetic Energy Inequality
 Leading to Lieb's Negative Ionization Upper Bound''
{\em Lett. Math. Phys.}  {\bf 28}, 219-230 (1993).

\bibitem{IM} K. Ito and H.P. McKean {\em Diffusion Processes and
Their Sample Paths} (Springer-Verlag, 1965).

\bibitem{LbRMP} E.H. Lieb, ``Thomas-Fermi and Related Theories
of Atoms and Molecules'' {\em Rev. Mod. Phys.} {\bf 53}, 603-641 (1981)
in \cite{LbSel}.

\bibitem{Lb}
E.H. Lieb, ``Bound on the Maximum Negative Ionization of Atoms and 
Molecules'' {\em Phys. Rev. A} {\bf 29}, 3018-328 (1984) in \cite{LbSel}.

\bibitem{LbSel}
E.H. Lieb, {\em The Stability of Matter: From Atoms to Stars}
Selecta  of E. Lieb, ed. by W. Thirring (second edition, Springer-Verlag, 1997).

\bibitem{LSST}
E.H. Lieb, I.M. Sigal, B. Simon, and W. Thirring, 
``Asymptotic Neutrality of Large-Z Ions''
{\em Commun. Math. Phys.} {\bf 116}, 635-644 (1988) 
 in \cite{LbSel}.
 
\bibitem{LSY1} E.H. Lieb, J.P. Solovej and J. Yngvason, 
``Heavy Atoms in the Strong Magnetic Field of a Neutron Star''
{\em Phys. Rev. Lett.} {\bf 69}, 749-752 (1992).

\bibitem{LSY2}  E.H. Lieb, J.P. Solovej and J. Yngvason, 
``Asymptotics of Heavy Atoms in High Magnetic Fields I:
Lowest Landau Band Regions'' 
{\em Commun. Pure Appl. Math.} {\bf 47}, 513-591 (1993).

\bibitem{LSY3} E.H. Lieb, J.P. Solovej and J. Yngvason, 
``Asymptotics of Heavy Atoms in High Magnetic Fields II:
Semiclassical Regions'' 
{\em Commun. Math. Phys.} {\bf 161}, 77-124 (1994). 

\bibitem{RMS} A.R.P. Rau, R.O. Mueller, and L. Spruch
``Simple Model and Wave Functions for Atoms in Intense Magnetic Fields''
{\em Phys. Rev. A} {\bf 11}, 1865-1879 (1975).

\bibitem{Rud} M. Ruderman, ``Matter in Superstrong Magnetic
Fields'' {\em Physics of Dense Matter}  C.J. Hansen, ed., pp. 117-131
(Rediel, Dordrecht, 1974).

\bibitem{Rusk1} M.B. Ruskai, 
`` Absence of Discrete Spectrum in Highly Negative Ions'' 
{\em Commun. Math. Phys.}  {\bf 82}, 457-469;  {\bf 85}, 325-327  (1982).

\bibitem{Rusk2} M.B. Ruskai,  
``Improved Estimates on the Number of Bound States of Negative
Bosonic Atoms'' {\em Ann. Inst. H. Poincare A: Physique Theorique}
{\bf  61}, 153-162 (1994).

\bibitem{RuW} M.B. Ruskai and E. Werner, 
``A Pair of Optimal Inequalities Related to the Error Function''
preprint (Austin Math. Phys. Archive 97-564).

\bibitem{SS} L. Schiff and H. Snyder, 
``Theory of the Quadratic Zeeman Effect''
{\em Phys. Rev.} {\bf 55}, 59-63 (1939).

\bibitem{SSS}  L.A. Seco, I.M. Sigal, P. Solovej
``Bound on the Ionization Energy of Large Atoms''
{\em Commun. Math. Phys.} {\bf 131}, 307-315 (1990).

\bibitem{Sig} I.M. Sigal, ``How Many Electrons Can a Nucleus Bind?''
{\em Ann. Phys.} {\bf 157}, 307-320 (1984).

\bibitem{SzW} S. J. Szarek and E. Werner, ``Confidence Regions for Means
of Multivariate Normal Distributions and a Non-symmetric Correlation
Inequality for Gaussian Measure'' 
{\em J. Multivariate Analysis} (in press).

\bibitem{W} M. Wirth, ``On consid\`ere la fonction de ${\mathbb R}$ dans 
${\mathbb R}$ d\'efine par $f(x) = e^{-x^2/2}$; d\'emontrer que la
fonction $g$ de ${\mathbb R}$ dans ${\mathbb R}$ d\'efine par 
$g(x) = f(x)/\int_x^\infty f(t) dt$ est convexe''
{\em Revue de Math\'ematiques Sp\'eciales} {\bf 104}, 187-88 (1993).

\bibitem{Y} J. Yngvason, 
``Asymptotics of Natural and Artificial Atoms in Strong Magnetic Fields''
{\em Proceeding of the XIth International Congress of Mathematical
Physics} D. Iagolnitzer, ed.,  pp. 185-205 (International Press,
Cambridge, Massachusetts, 1995) in \cite{LbSel}.

 

\end{thebibliography}
 \end{document}